\documentclass{scrartcl}
\usepackage[numbers,sort&compress]{natbib}
\usepackage{color}
\usepackage{bbm,amsmath}
\usepackage{graphicx}
\usepackage{amssymb}
\usepackage{authblk}

\usepackage{lineno}
\usepackage{nameref,hyperref}

\title{
Most undirected random graphs are amplifiers of selection for Birth-death dynamics,
but suppressors of selection for death-Birth dynamics
}
\author{
Laura Hindersin and Arne Traulsen}
\affil{
Department of Evolutionary Theory, Max Planck Institute for Evolutionary Biology, D-24306 Pl\"on, Germany.\\
}
\begin{document}

\maketitle
\section*{Abstract}
We analyze evolutionary dynamics on graphs, where the nodes represent individuals of a population.
The links of a node describe which other individuals can be displaced by the offspring of the individual on that node. 
Amplifiers of selection are graphs for which the fixation probability is increased for advantageous mutants and decreased for disadvantageous mutants. 
A few examples of such amplifiers have been developed, but so far it is unclear how many such structures exist and how to construct them. 
Here, we show that almost any undirected random graph is an amplifier of selection for Birth-death updating, where an individual is selected to reproduce with probability proportional to its fitness and one of its neighbors is replaced by that offspring at random.
If we instead focus on death-Birth updating, in which a random individual is removed and its neighbors compete for the empty spot, then the same ensemble of graphs consists of almost only suppressors of selection for which the fixation probability is decreased for advantageous mutants and increased for disadvantageous mutants. 
Thus, the impact of population structure on evolutionary dynamics is a subtle issue that will depend on seemingly minor details of the underlying evolutionary process.

\newpage

\section*{Introduction}
In the past years, it has been shown that evolution can be strongly affected by population structure, extending the classical result that ``certain quantities are independent of the geographical structure of a population''~\cite{maruyama:TPB:1974}.  
An approach that became popular in the past decade is to consider population structures described in terms of graphs, where the nodes are individuals which reproduce (or interact) through the links of the graph~\cite{lieberman:Nature:2005,nowak:book:2006,broom:PRSA:2008,broom:book:2013,maciejewski:JTB:2014}. 
It has been shown that all regular graphs do not change the probability that a mutant will either take over or go extinct compared to well-mixed populations, which can be described by a fully connected graph~\cite{lieberman:Nature:2005,nowak:book:2006,maruyama:TPB:1974}. 
This result has been obtained for a microscopic evolutionary process, called Birth-death (Bd) update, in which first an individual is sampled from the whole population at random, but proportional to fitness, and then its identical offspring replaces a neighboring individual.
However, such results depend on the details of the microscopic evolutionary process~\cite{maciejewski:JTB:2014,kaveh:JRSOS:2015}, which makes it challenging to disentangle structure from dynamics. 
Also the temporal dynamics of this process has interesting aspects, as amplifiers of selection may slow down the process of fixation itself~\cite{broom:PRSA:2010,frean:PRSB:2013,hindersin:JRSI:2014}.

Of particular interest are two specific classes of graphs, namely amplifiers and suppressors of selection.
A graph is called an \textit{amplifier of selection}, 
if, compared to a well-mixed population, 
(i) advantageous mutants have a higher fixation probability on this graph and
(ii) disadvantageous mutants have a lower fixation probability on this graph. 
Effectively, selection is amplified as mutants on this graph only require a lower fitness difference compared to the wild-type to obtain the same fixation probability as in the well-mixed population and consequently, drift is suppressed.
Some examples for amplifiers of selection are given in~\cite{lieberman:Nature:2005},
e.g.\ the star, the super-star or the funnel. 
  
Conversely, a graph is called a \textit{suppressor of selection}, 
if, compared to a well-mixed population, 
(i) advantageous mutants have a lower fixation probability on this graph and
(ii) disadvantageous mutants have a higher fixation probability on this graph. 
In~\cite{lieberman:Nature:2005}, examples for suppressors of selection are mostly source and sink populations, 
such as one-rooted graphs or hierarchical tissues~\cite{nowak:PNAS:2003,werner:PlosCB:2011}. 

The present study has been triggered by two observations: 
First, the star graph, which is a popular and simple amplifier of selection under Bd updating~\cite{lieberman:Nature:2005,broom:PRSA:2008,frean:PRSB:2013}, 
becomes a suppressor of selection under dB updating,
where a random individual is removed and its neighbors compete for the empty spot proportional to their fitness \cite{baxter:unpublished:2008}.
Second, for small populations under Bd updating, all possible undirected graphs are amplifiers of selection, unless they are regular and thus identical to the well-mixed population in terms of the fixation probability~\cite{hindersin:JRSI:2014}. 

The complexity of the amplifiers of selection and the simplicity of the suppressors of
selection given in~\cite{lieberman:Nature:2005} has suggested that it could be easier to
construct suppressors of selection than amplifiers of selection. 
In fact, it seems typically trivial to construct suppressors of selection (as one usually focuses on directed graphs to achieve this), whereas to construct amplifiers of selection seems a much more challenging task (as indicated by the sophistication of structures as the super-star or the meta-funnel). 
Does this imply that most population structures do suppress selection?

Here, we show that this strongly depends on the update mechanism. 
For Birth-death update with a mutant at a random initial position, 
it turns out that almost all small random graphs are actually amplifiers of selection.
For death-Birth updating, almost all small random graphs turn out to be suppressors of selection. 
This shows that it is basically trivial to construct either amplifiers or suppressors of selection, if the microscopic evolutionary process is not fixed. However, as most graphs have fixation probabilities close to the well-mixed case~\cite{adlam:SciRep:2014},
these effects may not be particularly strong and it may be much more challenging to construct strong amplifiers of selection. 
Another interesting observation is that the vast majority of connected random graphs are either amplifiers or suppressors. It seems to be more challenging to construct a graph that decreases the fixation probability for both advantageous and disadvantageous mutants, but we show that also for this case, a very simple example exists.

\section*{Methods}
We study the discrete-time Moran process on graphs, as discussed in~\cite{lieberman:Nature:2005}.
Each node of the graph represents an individual.
The outgoing links of a node describe where the offspring of the focal individual can be placed.
The incoming links of a node describe which other individuals can place their offspring into the focal node. 
We restrict ourselves to two types of individuals, wild-type individuals with fitness $1$ and mutant individuals with fitness $r$.
For $0<r<1$, mutants are disadvantageous, 
for $r>1$ they are advantageous.  
We start from a population of wild-types with a single mutant individual located on a random node of the graph. 
At each time step, there is one birth and one death event.
We compare two widely used update mechanisms. 
In the Birth-death update (Bd), 
an individual is chosen from the whole population with a probability proportional to its fitness. 
It reproduces and places its offspring randomly in one of its neighboring nodes.
Thus, selection is global. 
In the death-Birth update (dB), a random individual is chosen to die. 
One of its neighbors is selected with a probability proportional to its fitness to reproduce into the vacant node.
Thus, selection is local.

We do not analyze the case of selection at death (bD, Db) or selection at birth and death (BD, DB)~\cite{antal:PRL:2006,altrock:PRE:2009,debarre:NatComm:2014,kaveh:JRSOS:2015}.
We first focus on undirected graphs, where all links are bi-directional. 
For Bd, the outgoing link weights are given by $1/k_i$ where $k_i $ is the number of neighbors of node $i$. 
For dB, this normalization is implemented over the incoming links, such that their sum is $1$. 
This ensures that replacement is equally probable among the neighbors.

\subsection*{Fixation probabilities in well-mixed populations}

In the case of a well-mixed population, 
which corresponds to a complete graph, 
we can analytically calculate the fixation probability, 
i.e.\ the chance that a single mutant takes over the entire population.
This function, which should always increase with the mutant fitness $r$, is given for any Markov process with a tri-diagonal transition matrix and two absorbing states at $0$ and $N$ by~\cite{karlin:book:1975,nowak:book:2006,traulsen:bookchapter:2009}
\begin{equation}
\phi^M = \frac{1}{1+\sum_{k=1}^{N-1}{\prod_{j=1}^k{\frac{T^{j-}}{T^{j+}} }}},
\label{eq:fixprob}
\end{equation}
where $T^{j \pm}$ is the probability to change the number of mutants from $j$ to $j\pm1$.

In the case of the Bd update, we have
\begin{equation}
T^{j+}_{Bd} = \frac{j r}{j r + N-j} \frac{N-j}{N-1}
\quad \mathrm{and} \quad
T^{j-}_{Bd} = \frac{N-j}{j r + N-j} \frac{j}{N-1},
\label{eq:Bd_trans}
\end{equation}
where the $N-1$ implies that we have excluded the possibility that an offspring replaces its parent instead of a neighbor. 
For the fixation probability of a single mutant in the Bd process, we find
\begin{equation}
\phi_{Bd}^M 
= \frac{1-\frac{1}{r}}{1-\frac{1}{r^N}}. 
\label{eq:Bdprob}
\end{equation}
If we allowed for self-replacement, we would need to replace the $N-1$ in
Eq.~\eqref{eq:Bd_trans} by $N$, but we would recover the same result, as the ratio 
${T^{j-}_{Bd}}/{T^{j+}_{Bd}}= 1/r$ remains the same.
Thus, in well-mixed populations the difference is immaterial. 
In graph-structured populations, where self-replacement would imply the inclusion
of self-loops at all nodes, this issue can be more intricate. 
For Bd without self-replacement, it has been shown that for several random graph models the fixation probability of the random graphs converges to Eq.~\eqref{eq:Bdprob} with increasing graph size~\cite{adlam:SciRep:2014}.

In the case of the dB update, we find instead
\begin{equation}
T^{j+}_{dB} = \frac{N-j}{N} \frac{j r}{j r + N-j-1}
\quad \mathrm{and} \quad
T^{j-}_{dB} = \frac{j}{N} \frac{N-j}{(j-1) r + N-j},
\label{eq:dB_trans}
\end{equation}
where we have again excluded the possibility of self-replacement.
The fixation probability is now~\cite{kaveh:JRSOS:2015}
\begin{equation}
\phi_{dB}^M = \frac{1}{1+\sum_{k=1}^{N-1}{\prod_{j=1}^k{\frac{1}{r} \frac{j r + N - j - 1}{(j-1)r + N-j} }}}
= \frac{N-1}{N} \frac{1-\frac{1}{r}}{1-\frac{1}{r^{N-1}}}.
\label{eq:dBprob}
\end{equation}
In the case of the dB update, the fixation probability is different if we allow for self-replacement. 
However, for dB it would imply that the removed individual is competing with its neighbors for its own slot.
Therefore, the dB rule with self-replacement seems illogical when we think of the death of individuals.
It has however been used in a game-theoretical context as the so-called imitation updating \cite{ohtsuki:Nature:2006}, where a random individual is chosen to imitate the strategy of one of its neighbors or stick to its own strategy with a probability proportional to fitness. 
This update mechanism is obtained from the dB update by adding self-loops to every node of the graph.

\begin{figure}
	\centering
   	\includegraphics[width=\textwidth]{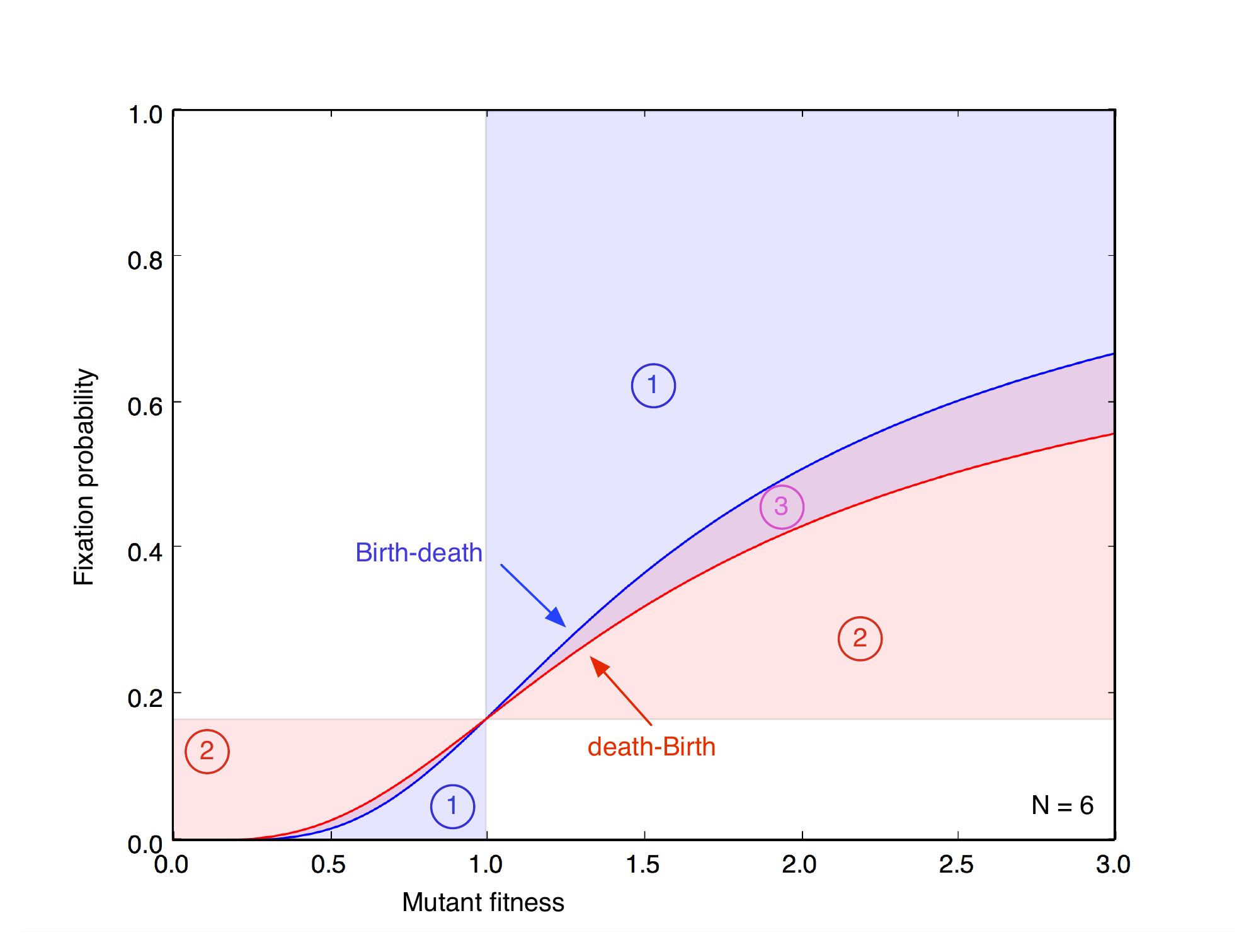}
\caption{\textbf{Comparison of Bd and dB update in the well-mixed population.}
Even in a well-mixed population, Birth-death and death-Birth processes do 
not lead to the same fixation probability, cf.\ Eqs.~\eqref{eq:Bdprob} and~\eqref{eq:dBprob}.  
This implies that we have to choose a different reference case in order to infer whether graph-structured populations are suppressors or amplifiers of selection. 
All fixation probabilities are expected to pass through the shaded region. 
For Bd updating, the fixation probability of amplifiers passes through region 1, whereas for dB updating, it passes through 1 or 3.
The fixation probability of suppressors for dB passes through 2, 
whereas for Bd it passes through 2 or 3.
Interestingly, when we directly compare the two well-mixed update processes, dB is a suppressor of selection compared to Bd -- and Bd is an amplifier of selection compared to dB.
}
\label{fig1}
\end{figure}

In the following, we exclude self-replacement and analyze 
how the update mechanisms change the fixation probability in graph-structured populations.
However, when we move to graphs, we should choose the corresponding well-mixed
case as a reference.
The difference between Eqs.~\eqref{eq:Bdprob} and~\eqref{eq:dBprob}, which becomes very small for large $N$, implies that the reference is not the same for the two update mechanisms, see Fig \ref{fig1}. 
For analytical approaches, which are based on the approximation of large $N$, this difference is not particularly important.
However, for us this difference can be crucial, as we are focusing on small graphs. 

\subsection*{Numerical Procedure}

We numerically generate a large number of Erd\H{o}s-R\'enyi graphs~\cite{erdos:PMHAS:1960} and calculate the fixation probability for a single mutant introduced at a random node on that graph.
The $G(N,p)$-algorithm~\cite{gilbert:ANMS:1959} generates a random graph with $N$ nodes where each link is independently present with probability $p$. 
We first analyze if the graph is connected, i.e.\ there are no isolated nodes. 
If it is disconnected, the fixation probability is formally zero and the graph is no longer of interest to our analysis.
In a next step, we check for isothermality~\cite{lieberman:Nature:2005}, i.e.\ whether every node has the same probability of being replaced.
If the graph is isothermal, the fixation probability in the Bd case is given by that of the complete graph~\cite{lieberman:Nature:2005}.
After creating the random graphs, we calculate the fixation probability for a single mutant in that respective graph, using the numerical approach described in~\cite{grinstead:book:1997,hindersin:JRSI:2014}. 
While it is tempting to use the symmetries of the graph to reduce the number of states, for our computational analysis it is sufficient to be able to number all states, i.e. if $\textbf{T}_{}$ is the transition matrix of this absorbing Markov chain, we reorder the states such that the $t$ transient states appear first and the two absorbing states, where all individuals are either mutants or wild-type, are last. 
Now the transition matrix is in the so-called canonical form,
\begin{equation}
\mathbf{T}_{(t+2) \times (t+2)}\ = \ \left(
 \begin{array}{cc}
  \mathbf{Q}& \mathbf{R}\\
 \mathbf{0}& \mathbf{I}\\
   \end{array} \right),
\end{equation}
where $\mathbf{Q}_{t\times t}$ contains the transition probabilities between transient states and $\mathbf{R}_{t\times 2}$ 
describes the transitions from the transient into the two absorbing states. 
Since the process can never move out of absorbing states, the lower left block of size $2 \times t$ is zero  and the lower right block is the identity matrix of size $2 \times 2$. 
Given a starting distribution $x_{1\times (t+2)}$, the product $x\mathbf{T}^m$ yields the distribution after exactly $m$ time steps.
For $m \to \infty$, we recover the fixation probabilities. 

Summing over the transient part $\mathbf{Q}$, we obtain the so-called fundamental matrix of the Markov chain $\mathbf{F} = \sum_{n=0}^{\infty}{\mathbf{Q}^n} = (\mathbf{I}-\mathbf{Q})^{-1} $. 
The $(i,j)$-th entry of $\mathbf{F}$, 
\begin{equation}
F_{i,j} = \left( \left( \mathbf{I}-\mathbf{Q} \right)^{-1} \right)_{i,j}
\end{equation} 
is the expected sojourn time in the transient state $j$, given that the process started in the transient state $i$. 
Multiplying $\mathbf{F}$ with the transition probabilities to the absorbing states provides the fixation probabilities that are of interest to us here. 
The fixation probability in state $j$ after starting in state $i$,
$\phi_{i,j}$, is the $(i,j)$-th entry of $\Phi= \mathbf{F} \cdot \mathbf{R}$, 
\begin{equation}
\phi_{i,j} = \left( \left( \mathbf{I}-\mathbf{Q} \right)^{-1} \mathbf{R} \right)_{i,j}.
\end{equation}
Our main quantity of interest is the probability of a single mutant, introduced at a random node, taking over a population of wild-type individuals.

This fixation probability, which we denote by $\phi^G$ on a graph $G$, is a function of the fitness $r$. 
We do not solve the linear system analytically for general $r$, as this would lead to rational functions of high degrees which are difficult to interpret. 
Instead, we focus on the five numerical values $r=0.75,1,1.25,1.5,1.75$. 
This is enough to classify most graphs as either an amplifier or a suppressor of selection. 
Our classification is as follows:
\begin{itemize}
\item If $\phi^G < \phi^M$ for $r=0.75$ and $\phi^G > \phi^M$ for $r=1.25,1.5,1.75$ \quad $\Longrightarrow$ \quad Amplifier
\item If $\phi^G > \phi^M$ for $r=0.75$ and $\phi^G < \phi^M$ for $r=1.25,1.5,1.75$ \quad $\Longrightarrow$ \quad Suppressor
\end{itemize}
If neither of these two conditions is true, we call the graph ``unclassified''.
This can either happen due to small numerical errors in our implementation in Python
or because the graph is truly neither an amplifier nor a suppressor of selection 
(see Eq.~\eqref{eq:ring-mix} for an example).

This numerical approach is limited due to the system size of up to 
$2^N$ states and accordingly a transition matrix of size up to $2^N \times 2^N$. 
Taking symmetries into account could substantially decrease this number, but this
would need to be done on a case by case basis -- an approach that is not suitable if
we focus on a large number of random graphs. 
Alternatively, we could perform stochastic simulations of the fixation process
and obtain the fixation probabilities from averaging over many realizations. 
However, as the fixation probabilities on graphs are typically close to neutral~\cite{adlam:SciRep:2014}, the precision necessary to classify our graphs 
requires a very large number of realizations for each $r$. 
In contrast, our numerical approach described above has to be performed only once for each value of $r$ and does not suffer from any noise which would arise from averaging over a stochastic process.

\section*{Results}
\subsection*{Birth-death}

In~\cite{hindersin:JRSI:2014}, we have shown that all four undirected, degree-heterogeneous graphs of size $4$ are amplifiers of selection. 
The remaining two graphs, the cycle and the complete graph, are degree-homogeneous and thus have the same fixation probability as the well-mixed population. 
How does this change when we move to larger graphs? 
The number of possible graphs is $2^{N(N-1)/2}$, i.e.\ it increases rapidly with $N$. 
This combinatorial explosion prevents us from analyzing all graphs systematically. 
Instead, we apply the procedure described above and focus on random graphs, using the case of $N=4$ to cross-check.

Let $P_N$ be the probability that a graph with $N$ nodes is connected. 
This probability is given recursively by~\cite{gilbert:ANMS:1959}
\begin{equation}
P_N = 1 - \sum_{K=1}^{N-1} {\binom{N-1}{K-1} P_K (1-p)^{K(N-K)}},
\label{eq:ProbConnected}
\end{equation}
where $P_1=1$.
For example, the probability that a graph with $N=4$ nodes is connected is $P_4 = -6p^6 + 24p^5 - 33 p^4 +16p^3$.

\begin{figure}
	\centering
   	\includegraphics[width=\textwidth]{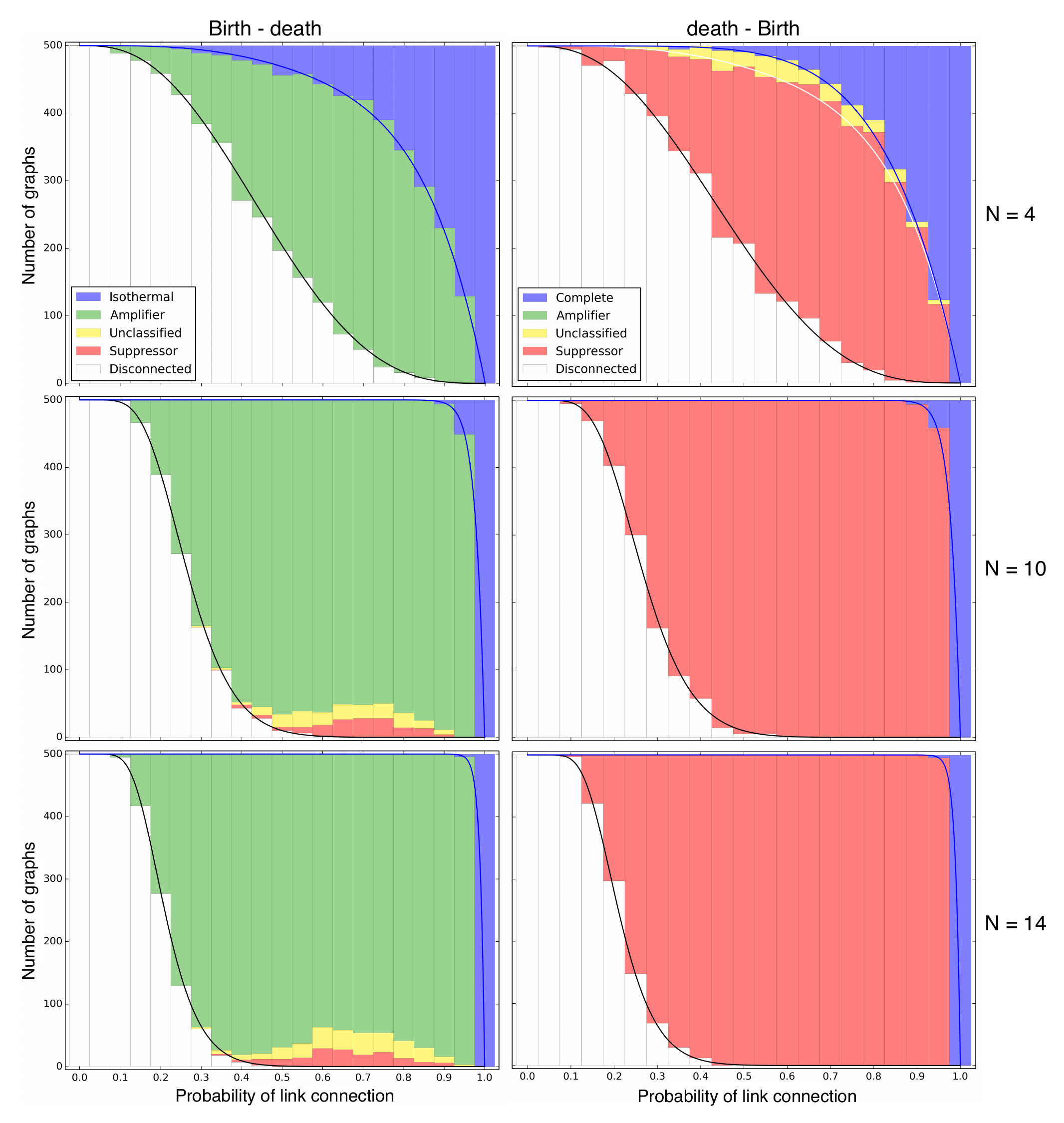}
\caption{\textbf{Bd and dB update on undirected random graphs.}
Fixation probability for the Moran process on undirected random graphs for varying probability of link connection $p$ in Erd\H{o}s-R\'enyi graphs. 
The black line depicts the proportion of connected graphs, given by equation \eqref{eq:ProbConnected}. 
In the left panels, the blue line yields the proportion of isothermal graphs.
Since the isothermal theorem does not hold for dB \cite{kaveh:JRSOS:2015}, in the right panels the blue line depicts only the complete graphs, whereas the white line gives the proportion of cycles for $N=4$. 
Left: For Birth-death updating, most non-trivial graphs are amplifiers of selection and only a tiny fraction are either suppressors of selection or remain unclassified.  
Right: If the population size is sufficiently large for death-Birth updating, we only find suppressors of selection. 
The only ``unclassified" graph of size $N=4$ is the cycle, which is neither an amplifier nor a suppressor of selection for dB, see Eq.~\eqref{eq:ring-mix}.
}
\label{fig2}
\end{figure}

The probability that a graph is complete is $p^{N(N-1)/2}$.
The probability that a graph of size $N=4$ is a cycle is 
$3 p^4(1-p)^2$.
Thus, the probability that a graph of size $N=4$ is isothermal is
$p^6+3p^4(1-p)^2$, see Fig \ref{fig2}.

In general, most graphs are disconnected for small $p$. 
For $p$ approaching $1$, most graphs are just the complete graph. 
Fig \ref{fig2}. shows that for Bd updating, the vast majority of graphs are amplifiers of selection. 
For larger graphs, there are some graphs that are suppressors of selection
and some that remain unclassified. 

Only for $p$ very close to 1 there are isothermal graphs, because  e.g.\ for size 10 it is already very unlikely that a $k$-regular graph of degree $k<N-1$ like the cycle is constructed by chance. 
Thus, the proportion of isothermal graphs approaches $p^{N(N-1)/2}$ when $N$ becomes large.

\subsection*{death-Birth}

The star, a popular amplifier of selection under Bd updating, is actually a 
suppressor of selection for dB updating~\cite{baxter:unpublished:2008}.
It turns out that a very large proportion of graphs are suppressors of selection, see Fig \ref{fig2}. 
Up to size 14, we have not found a single amplifier with our procedure, suggesting that they
are extremely rare or do not exist at all. 

\begin{figure}
	\centering
   	\includegraphics[width=\textwidth]{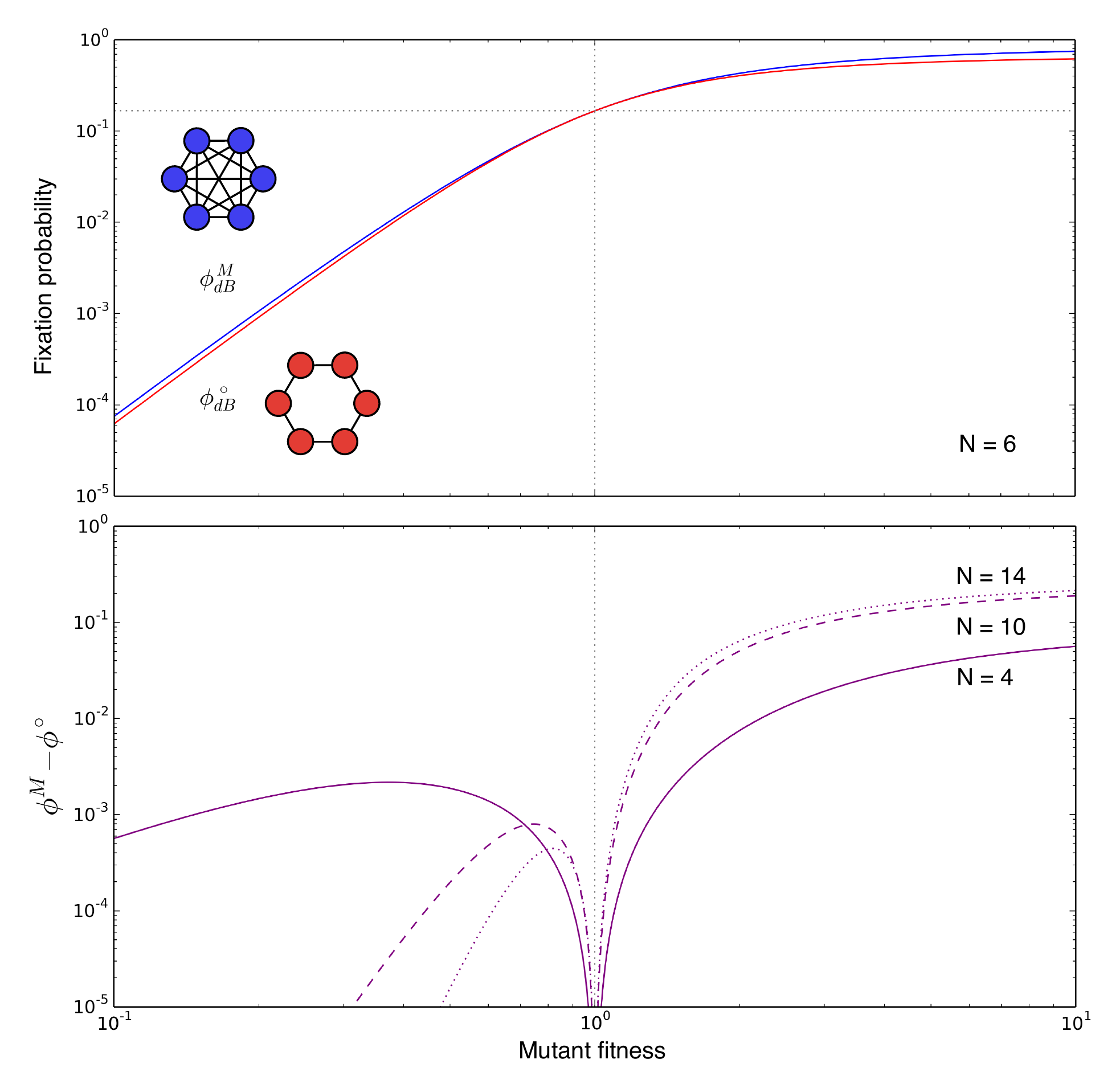}
\caption{\textbf{Comparison of the cycle and the well-mixed population.}
Top: The cycle for small $N$ is neither an amplifier nor a suppressor of selection. 
It decreases the fixation probability compared to the well-mixed population for both advantageous and disadvantageous mutants.	
Bottom: The difference between the fixation probability in the well-mixed population and the cycle increases with $N$ for advantageous mutants. 
For disadvantageous mutants, all fixation probabilities tend to zero as the graph size increases.
}
\label{fig3}
\end{figure}

For $N=4$, we find one graph that is neither an amplifier of selection nor a suppressor of selection. 
It turns out that this is the cycle, which occurs with probability $3 p^4(1-p)^2$.
It has been shown recently \cite{kaveh:JRSOS:2015} that the isothermal theorem does not hold for dB updating.
For the cycle of size $N=4$, we can calculate the fixation probability of a randomly placed mutant analytically. 
For the transition probabilities, we have 
\begin{align}
& T^{1+}  =\ \frac{1}{2} \frac{r}{r+1} 	& T^{1-}  &=\ \frac{1}{4} \\  
& T^{2+}  =\ \frac{1}{2} \frac{r}{r+1} 	& T^{2-} &=\ \frac{1}{2} \frac{1}{r+1} \\  
& T^{3+}  =\ \frac{1}{4}			& T^{3-}  &=\ \frac{1}{2} \frac{1}{r+1}  \ .
\end{align}
With Eq.~\eqref{eq:fixprob}, this leads to 
\begin{equation}
\phi_{dB }^{\circ}= \frac{2r^2}{3r^2+2r+3} \ .
\end{equation}
Comparing this to the corresponding well-mixed case $\phi^M_{dB}$ from Eq.~\eqref{eq:dBprob}, we have

\begin{align}
\begin{split} 
\label{eq:ring-mix}
\phi_{dB}^{\circ} - \phi_{dB}^M  \quad  &= \quad \frac{2r^2}{3r^2+2r+3} - \frac{3r^2}{4(r^2+r+1)}   \\
\quad &= \quad - \ \frac{r^2 (r-1)^2}{4(r^2+r+1)(3r^2+2r+3)} \ .
\end{split}
\end{align}
Obviously, we have $\phi_{dB}^{\circ} < \phi_{dB}^M$ for any $r>0$, $r \neq 1$, which implies
that the fixation probability on the cycle with dB updating is smaller than the corresponding
well-mixed population both for disadvantageous and advantageous mutations. 
Thus, the cycle with dB updating is neither an amplifier nor a suppressor of selection (see Fig \ref{fig3}.).
In the Supporting Information, 
we show that this holds for general $N$, i.e.\
\begin{equation}
\phi_{dB}^M > \phi_{dB}^{\circ}  \quad \quad \textrm{for} \  r>0 
\label{eq:generalN}
\end{equation}
and $\phi_{dB}^M = \phi_{dB}^{\circ}$ for $r=1$. 
In other words, the cycle with dB updating is a suppressor of fixation 
compared to the corresponding well mixed case. 

\subsection*{Directed graphs}
The examples of suppressors of selection for Bd updating given in \cite{lieberman:Nature:2005,nowak:book:2006} are directed graphs, e.g.\ the directed line with a fixation probability of $1/N$, meaning that selection is completely eliminated.
Another example are so-called source and sink graphs, where fixation in the upstream population is sufficient for a fixation in the whole graph, which suppresses selection.
To explore the abundance of amplifiers and suppressors of selection on directed graphs, we use the numerical procedure explained in the Methods section. In addition, we need to check whether a graph is rooted.
If a graph of size $N$ is one-rooted, meaning that there is exactly one node with in-degree zero and positive out-degree, then the fixation probability is $1/N$ because the newly introduced mutant can only reach fixation if it is placed on that root node.
On a multi-rooted graph, the mutant can never fixate, therefore the fixation probability is zero.

\begin{figure}
	\centering
   	\includegraphics[width=\textwidth]{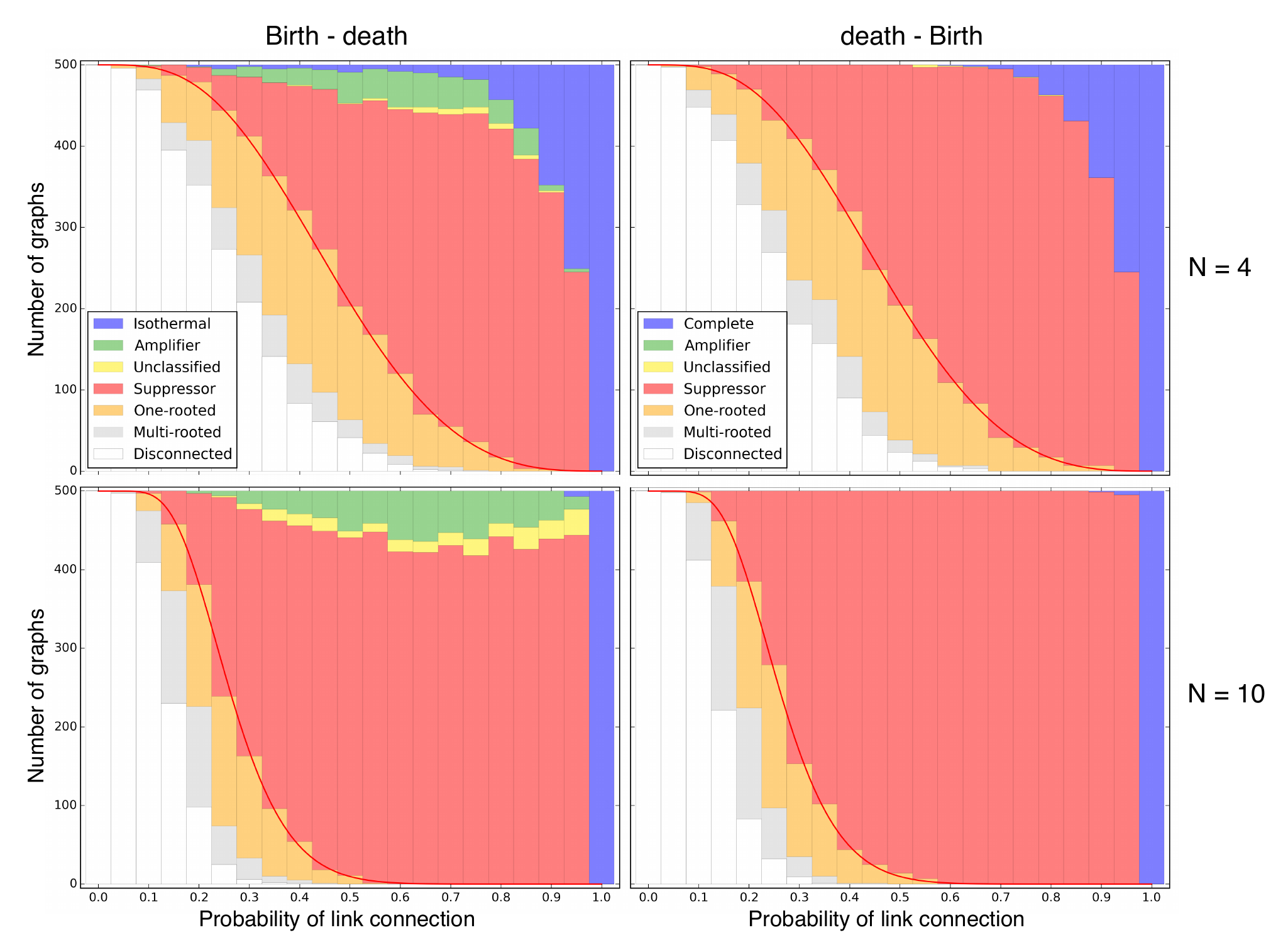}
\caption{\textbf{Bd and dB update on directed random graphs.}
500 directed random graphs are classified into amplifiers or suppressors of selection according to their fixation probability for one randomly placed mutant.
The red line represents the proportion of graphs having at least one node with no incoming links, see Eq.~\eqref{eq:rootedGraphs}, which is equivalent to the sum of the disconnected, multi-rooted and one-rooted graphs.
Technically, one-rooted graphs are also suppressors of selection, because a randomly placed mutant has a fixation probability of $1/N$ in such a graph.
We visualize them in orange to distinguish them from the other suppressors.
Multi-rooted graphs, given in grey, have a fixation probability of zero.
The seven ``unclassified" graphs we found for dB updating of size $N=4$ are the same as the undirected cycle, in this case given by eight directed links instead of four undirected links.
As shown above, the cycle is neither an amplifier nor a suppressor of selection.
}
\label{fig4}
\end{figure}

In a directed Erd\H{o}s-R\'enyi random graph constructed by the $G(n,p)$-algorithm, each link is independently present with probability $p$.
Thus, the probability that a node has in-degree zero is $(1-p)^{N-1}$, as we exclude self-loops.
Therefore the probability that there is at least one node with in-degree zero is given by
\begin{equation}
1 - ( 1 - (1-p)^{N-1} )^N .
\label{eq:rootedGraphs}
\end{equation}
Fig \ref{fig4}. displays 500 directed random graphs for Bd and dB updating. 
For dB there are no amplifiers, but almost only suppressors, exactly as in the undirected case.
For Bd, we still find some amplifiers, but also many suppressors of selection, in contrast to the undirected case.

\section*{Discussion}
We have shown that small random graphs are mostly either amplifiers or suppressors of selection, depending on the microscopic evolutionary process at stake. 
It is interesting that very few structures fall outside this classification, with the cycle under dB updating as a simple example. 
Naively, one would expect many more possibilities beyond amplifiers or suppressors of selection.
However, our numerical approach has the potential shortcoming that we cannot entirely exclude that such cases are misclassified, as we focus on a relatively small number of values for $r$ to classify the graphs. 
For example, the fixation probability could in principle have different asymptotics for large $r$ and it could intersect with the well-mixed case \cite{voorhees:PRSA:2013a}.  
However, the same shortcoming would exist if we instead perform stochastic simulations.

Both computational approaches to address the problem have very different characteristics. 
The numerical approach is based on the analysis of very large transition matrices
(in our case of $N=14$, we consider a matrix of $\approx 2.7 \cdot 10^8$ entries).
So increasing $N$ requires substantial memory and is additionally limited by the chosen numerical precision. 
On the other hand, it only needs a single computation for each value of $r$. 
The simulation based approach requires only a minimum of memory and can easily be parallelized on a computer cluster. 
But as the fixation probabilities are typically very close to the neutral case, a huge number of averages is necessary to reliably distinguish weak amplifiers or suppressors of selection from the corresponding reference case. 
An analytical procedure would clearly be desirable, but previous authors had to focus on certain approximations or rely on strong assumptions such as highly symmetrical graphs~\cite{houchmandzadeh:NJP:2011,monk:PRSA:2014,adlam:SciRep:2014,kaveh:JRSOS:2015,diaz:PRSA:2013,jamieson-lane:JTB:2015}.
Thus, a numerical exploration of this issue, as presented here, may be helpful 
to assess the relative proportion of the different graph classes first 
and to find illustrative examples that go beyond the classes of amplifiers and suppressors. 

The issue of update mechanisms has been analyzed in great depth in the context of evolutionary game theory, where the individuals not only reproduce via the links but also interact through them, leading to a fitness that depends on the local neighborhood~\cite{nowak:Nature:1992b,hauert:IJBC:2002,santos:PNAS:2006,szabo:PR:2007,tomassini:IJMP:2007,roca:PLR:2009,perc:BioSys:2010,perc:JRSI:2013,allen:EMS:2014}. 
In this case, it is particularly interesting that the Bd update cannot promote cooperation, whereas the dB update can~\cite{ohtsuki:Nature:2006,ohtsuki:PRSB:2006,taylor:Nature:2007,grafen:JEB:2007}.
Intuitively, the reason for this is that the initial growth of a disadvantageous mutant is much more likely for dB than for Bd. 
A careful comparison based on performing each of these updates with a certain probability shows that the case of only Bd updating is special, with the result resembling the pure dB updating as soon as a minimum of dB updating is used~\cite{zukewich:PlosOne:2013}.
Here, we have focused entirely on the pure cases to illustrate that even in the case of constant mutant fitness, which lacks the complexities of game theoretic interactions, 
substantial differences in the microscopic evolutionary processes change the way an ensemble of random graphs affects the fixation of mutants. 

An issue that we have not discussed here is where mutants would predominantly arise~\cite{santos:PNAS:2006,zukewich:PlosOne:2013,maciejewski:PlosCB:2014,allen:PlosCB:2015}. 
For example, on a star graph it may be much more likely that a mutant that occurs during reproduction is placed in the central node, simply because most reproducing individuals have only this node as a neighbor. This leads to additional complications, as the classification of suppressors and amplifiers of selection immediately depends on the microscopic evolutionary process and the mode of mutation. Instead of addressing this issue, \cite{maciejewski:PlosCB:2014,adlam:PRSA:2015}, we have concentrated on the fate of a single mutant that is located on a random initial node chosen with uniform probability. 

We have focused entirely on relatively small graphs, whereas most previous studies looked at much larger population sizes. 
Concentrating on these small graphs allows to establish clear classifications and a detailed understanding of particular cases such as the cycle. 
We expect that our results can be extended to much larger graphs and to different classes of graphs, but this will require substantial numerical efforts and possible novel simulation approaches.

To sum up, we have shown that it is trivial to construct weak amplifiers of selection: Almost all small undirected graphs fulfill these requirements under Birth-death updating. 
Interestingly, the same ensemble of random graphs consists entirely of suppressors of selection if we switch to death-Birth updating. 
The effect of population structure on evolution, which has by now been firmly established, remains a subtle issue.

\section*{Supporting Information}
\subsection*{Proof for Eq.~(15).}
\label{S1_Text}
\textbf{Here, we prove that a mutant on the cycle with dB updating has lower fixation probability than in the well-mixed population for general $N$.}
We have to show that $\phi_{dB}^M - \phi_{dB}^{\circ} > 0$ for $r>0, r\neq1$.
Under dB updating, the fixation probability in the well-mixed population is given by
\begin{equation}
\phi_{dB}^M =  \frac{N-1}{N} \frac{1-\frac{1}{r}}{1-\frac{1}{r^{N-1}}}.
\end{equation}
On the cycle, the respective fixation probability is given by (Eq.~(5.3) in \cite{kaveh:JRSOS:2015})
\begin{equation}
\phi_{dB}^{\circ} =  \frac{2(r-1)}{3r -1 + (r-3) r^{2-N}}.
\end{equation}
Then the difference is given by
\begin{equation}
\phi_{dB}^M - \phi_{dB}^{\circ} = \frac{(r-1) r^{N-2}}{N}
\cdot
\frac{ (N-3) r^{N+1} - (N-1) r^N + (N-1) r^3 - (N-3 ) r^2}
{ (r^{N-1} -1) (3 r^{N+1} - r^N + r^3 - 3 r^2)}
\end{equation}
This expression can be written in the form of
\begin{equation}
\frac{(r-1)^2 r^{N-2}}{N}
\cdot
\frac{ \sum_{k=0}^{N-4}{ r^k ((N-2)(k+1)-(k+1)^2)}  }
{\left( \sum_{k=0}^{N-2}{r^k} \right)  \left( 3 + \sum\limits_{k=1}^{N-3}{2 r^k} + 3 r^{N-2} \right)} \ .
\label{eq:diffEven}
\end{equation}
As Eq.~\eqref{eq:diffEven} contains only positive coefficients in $r$, the difference is always positive for $r > 0$ and $r\neq 1$.
For $r=1$, it is zero.
Thus  Eq.~(15) is fulfilled for all $r>0$, $r\neq 1$.

\section*{Acknowledgments}
We thank Kamran Kaveh, Benedikt Bauer, Bin Wu, and Jorge Pe\~{n}a for helpful discussions.



\end{document}